# On the Effective Capacity of Two-Hop Communication Systems

Deli Qiao, Mustafa Cenk Gursoy, and Senem Velipasalar
Department of Electrical Engineering
University of Nebraska-Lincoln, Lincoln, NE 68588
Email: dqiao726@huskers.unl.edu, gursoy@engr.unl.edu, velipasa@engr.unl.edu

***Abstract*—[1] In this paper, two-hop communication between a source and a destination with the aid of an intermediate relay node is considered. Both the source and intermediate relay node are assumed to operate under statistical quality of service (QoS) constraints imposed as limitations on the buffer overflow probabilities. It is further assumed that the nodes send the information at fixed power levels and have perfect channel side information. In this scenario, the maximum constant arrival rates that can be supported by this two-hop link are characterized by finding the effective capacity. Through this analysis, the impact upon the throughput of having buffer constraints at the source and intermediate-hop nodes is identified.**

## I. Introduction

Providing quality of service (QoS) guarantees is important in the design of the next generation wireless systems. For instance, in wireless systems that support voice over IP (VoIP) or streaming-video applications, the key QoS metric is delay. At the same time, satisfying these QoS considerations is challenging in wireless communication scenarios. Due to mobility, changing environment and multipath fading, the power of the received signal, and hence the instantaneous rates supported by the channel, fluctuate randomly [1]. In such a volatile environment, providing deterministic delay guarantees either is not possible or, when it is possible, requires the system to operate pessimistically and achieve low performance underutilizing the resources. Therefore, wireless systems are better suited to support statistical QoS guarantees.

In [2], Chang employed the effective bandwidth theory to analyze systems operating under statistical QoS constraints. These constraints are imposed on buffer violation probabilities and are specified by the QoS exponent $\theta$, which is defined as

$$\lim_{Q_{\text{max}}\to\infty} \frac{\log Pr\{Q(\infty) > Q_{\text{max}}\}}{Q_{\text{max}}} = -\theta, \tag{1}$$

where $Q(\infty)$ is the queue length in steady state, $Q(\text{max})$ is the maximal queue length. Therefore, QoS exponent $\theta$ is the exponential decay rate of the buffer overflow probability for large $Q(\text{max})$. A larger $\theta$ implies a lower probability of violating the queue length and is a more stringent QoS constraint. In [3], Chang and Zajic characterized the effective bandwidths of the time varying departure processes. In [4], Chang and Thomas applied the effective bandwidth theory to high-speed digital networks. More recently, Wu and Negi in [5] defined the dual concept of effective capacity, which provides the maximum constant arrival rate that can be supported by a given departure process while satisfying statistical delay constraints. Tang and Zhang in [6]-[8] employed the effective capacity formulation to conduct a performance analysis under different channel settings. For instance, the optimal power control policies that maximize the effective capacity of a point-to-point link have been derived in [6].

Although Chang in [3] provided an algorithm to obtain the effective bandwidths of different links for intree networks, no prior work has addressed the effective capacity of intree networks. Tang and Zhang in [8] analyzed the power allocation policies of relay networks, where the relay node is assumed to have no queue, i.e., the packets arriving to the relay node are decoded correctly and relayed immediately. Parag and Chamberland in [9] provided a queueing analysis of a butterfly network with constant rate for each link. However, they assumed that there is no congestion at the intermediate nodes.

In this paper, we attempt to characterize the maximum constant arrival rates that can be supported by a two-hop communication link when both the source and the intermediate relay node are operating under buffer constraints. The individual QoS constraints of the source and the intermediate nodes are described by the QoS exponents $\theta_1$ and $\theta_2$, respectively. We assume that the channel state knowledge of each link is known at both nodes, and the transmission power levels are fixed. Under these assumptions, we determine the effective capacity as a function of $\theta_1$ and $\theta_2$.

The rest of this paper is organized as follows. In Section II, the system model and necessary preliminaries are described. In Section III, we describe our our main result. Finally, in Section IV, we conclude the paper.

## II. System Model and Preliminaries

The two-hop communication link is depicted in Fig. 1. In this model, source $S$ is sending information to the destination $D$ with the help of the intermediate node $H$. Both the source and the intermediate relay node have QoS constraints specified by the QoS exponents $\theta_1$ and $\theta_2$.

[1]This work was supported by the National Science Foundation under Grants CNS–0834753, and CCF–0917265.

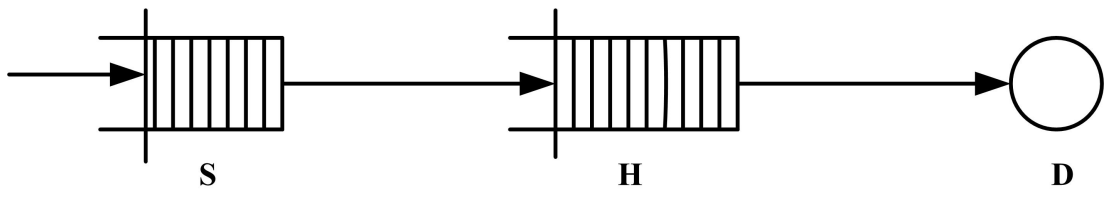

Fig. 1. The system model.

The discrete-time input and output relationships in the $i$th symbol duration are given by

$$Y[i] = g_1[i]X_1[i] + n_1[i] \quad (2)$$
$$Z[i] = g_2[i]X_2[i] + n_2[i] \quad (3)$$

where $X_j, j = \{1,2\}$ denote the inputs for the links $S-H$ and $H-D$, respectively. The inputs are subject to individual average energy constraints $\mathbb{E}\{|X_j|^2\} \le \bar{P}_j/B, j = \{1,2\}$ where $B$ is the bandwidth. We assume that the fading coefficients $g_j, j = \{1,2\}$ are jointly stationary and ergodic discrete-time processes, and we denote the magnitude-square of the fading coefficients by $z_j[i] = |g_j[i]|^2$. Assuming that there are $B$ complex symbols per second, we can easily see that the symbol energy constraint of $\bar{P}_j/B$ implies that the channel input has a power constraint of $\bar{P}_j$. Above, in the channel input-output relationships, the noise component $n_j[i]$ is a zero-mean, circularly symmetric, complex Gaussian random variable with variance $\mathbb{E}\{|n_j[i]|^2\} = N_j$ for $j = 1, 2$. The additive Gaussian noise samples $\{n_j[i]\}$ are assumed to form an independent and identically distributed (i.i.d.) sequence. We denote the signal-to-noise ratios as $\text{SNR}_j = \frac{\bar{P}_j}{N_j B}$. In order for the queues to be stable, we assume that $\text{SNR}_1$and $\text{SNR}_2$ are chosen such that $\mathbb{E}_{z_1}\{\log_2(1+\text{SNR}_1 z_1)\} < \mathbb{E}_{z_2}\{\log_2(1+\text{SNR}_2 z_2)\}$ [3].

We first state the following result from [3].

*Lemma 1 ([3]):* Suppose that the queue is stable and that both the arrival process $a[n], n = 1, 2, \ldots$ and service process $c[n], n = 1, 2, \ldots$ satisfy the Gärtner-Ellis limit, i.e., for all $\theta \ge 0$, there exists a differentiable logarithmic moment generating function (LMGF) $\Lambda_A(\theta)$ such that[2]

$$\lim_{n\to\infty} \frac{\log \mathbb{E}\{e^{\theta \sum_{i=1}^{n} a[n]}\}}{n} = \Lambda_A(\theta), \quad (4)$$

and a differentiable LMGF $\Lambda_C(\theta)$ such that

$$\lim_{n\to\infty} \frac{\log \mathbb{E}\{e^{\theta \sum_{i=1}^{n} c[n]}\}}{n} = \Lambda_C(\theta). \quad (5)$$

If there exists a unique $\theta^* > 0$ such that

$$\Lambda_A(\theta^*) + \Lambda_C(-\theta^*) = 0, \quad (6)$$

then

$$\lim_{Q_{\max}\to\infty} \frac{\log Pr\{Q(\infty) > Q_{\max}\}}{Q_{\max}} = -\theta^*. \quad (7)$$

□

Assume that the arrival rate to the transmitter is $R \ge 0$, and the channels operate at their capacities. To satisfy the QoS constraint at the source, we should have

$$\tilde{\theta} \ge \theta_1 \quad (8)$$

where $\tilde{\theta}$ is the solution to

$$R = -\frac{\Lambda_C(-\theta)}{\theta} \quad (9)$$

and $\Lambda_C(\theta)$ is the LMGF of the capacity of the $S-H$ link.

According to [3], the LMGF of the departure process from the transmitter, or equivalently the arrival process to the hop node, is given by

$$\Lambda_B(\theta) = \begin{cases} R\theta, & 0 \le \theta \le \tilde{\theta} \\ R\tilde{\theta} + \Lambda_C(\theta - \tilde{\theta}), & \theta > \tilde{\theta} \end{cases} \quad (10)$$

Therefore, in order to satisfy the QoS of the intermediate relay node $H$, we must have

$$\hat{\theta} \ge \theta_2 \quad (11)$$

where $\hat{\theta}$ is the solution to

$$\Lambda_B(\theta) + \Lambda_H(-\theta) = 0. \quad (12)$$

Above, $\Lambda_H(\theta)$ is the LMGF of the capacity of the $H-D$ link.

After these characterizations, effective capacity of the two-hop communication model can be formulated as follows.

*Definition 1:* The effective capacity of the two-hop communication link with the QoS constraints specified by $\theta_1$ at the source and $\theta_2$ at the relay node is given by

$$R_E(\theta_1, \theta_2) = \sup_{R\in\mathcal{R}} R \quad (13)$$

where $\mathcal{R}$ is the collection of arrival rates $R$ such that the solutions $\tilde{\theta}$ and $\hat{\theta}$ of (9) and (12) with any $R \in \mathcal{R}$ satisfy (8) and (11), respectively.

## III. Effective Capacity of a Two-Hop Link in Block Fading

We assume that the channel state information of the links $S-H$ and $H-D$ is available at $S$ and $H$, and the channel state information of the link $H-D$ is available at $D$. The transmission power levels at the source and the intermediate-hop node are fixed and hence no power control is employed. We further assume that the channel capacity for each link can be achieved, i.e., the service processes are equal to the instantaneous Shannon capacities of the links. We consider a block fading scenario in which the fading stays constant for a block of $T$ seconds and change independently from one block to another.

Under the block fading assumption, the logarithmic moment generating function for the service processes of links $S-H$ and $H-D$ as functions of $\theta$ are given by [6]

$$\Lambda_C(\theta) = \log \mathbb{E}_{z_1}\left\{e^{-\theta TB \log_2(1+\text{SNR}_1 z_1)}\right\} \quad (14)$$
$$\Lambda_H(\theta) = \log \mathbb{E}_{z_2}\left\{e^{-\theta TB \log_2(1+\text{SNR}_2 z_2)}\right\} \quad (15)$$

[2]Throughout the text, logarithm expressed without a base, i.e., $\log(\cdot)$, refers to the natural logarithm $\log_e(\cdot)$.

and as a result

$$\Lambda_B(\theta) = \begin{cases} R\theta, & 0 \le \theta \le \tilde{\theta} \\ R\tilde{\theta} + \log \mathbb{E}_{z_1}\left\{e^{-(\theta-\tilde{\theta})TB\log_2(1+\text{SNR}_1 z_1)}\right\}, & \theta > \tilde{\theta} \end{cases}$$

Below, we provide our main result.

*Theorem 1:* The effective capacity of the two-hop communication link with statistical QoS constraints at the source and the intermediate relay nodes specified by $\theta_1$ at the source and $\theta_2$ at the relay is given by

**Case I** $\theta_1 \ge \theta_2$ :

$$R_E(\theta_1,\theta_2) = \min\left\{ -\frac{1}{\theta_1}\log \mathbb{E}_{z_1}\left\{e^{-\theta_1 TB\log_2(1+\text{SNR}_1 z_1)}\right\}, -\frac{1}{\theta_2}\log \mathbb{E}_{z_2}\left\{e^{-\theta_2 TB\log_2(1+\text{SNR}_2 z_2)}\right\}\right\}$$

**Case II** $\theta_1 < \theta_2$ :

1) $\theta_2 \le \theta_2'$,

$$R_E(\theta_1,\theta_2) = -\frac{1}{\theta_1}\log \mathbb{E}_{z_1}\left\{e^{-\theta_1 TB\log_2(1+\text{SNR}_1 z_1)}\right\}$$

2) $\theta_2 > \theta_2'$,

$$R_E(\theta_1,\theta_2) = -\frac{1}{\tilde{\theta}_0}\log \mathbb{E}_{z_1}\left\{e^{-\tilde{\theta}_0 TB\log_2(1+\text{SNR}_1 z_1)}\right\}$$

where $\theta_2'$ is the solution to

$$\begin{aligned} &-\frac{1}{\theta_1}\log \mathbb{E}_{z_1}\left\{e^{-\theta_1 TB\log_2(1+\text{SNR}_1 z_1)}\right\}, \\ &= -\frac{1}{\theta_1}\Big(\log \mathbb{E}_{z_2}\left\{e^{-\theta TB\log_2(1+\text{SNR}_2 z_2)}\right\} \\ &\quad + \log \mathbb{E}_{z_1}\left\{e^{(\theta-\theta_1) TB\log_2(1+\text{SNR}_1 z_1)}\right\}\Big) \end{aligned} \tag{16}$$

and $\tilde{\theta}_0$ is the solution to

$$\begin{aligned} &-\frac{1}{\tilde{\theta}}\log \mathbb{E}_{z_1}\left\{e^{-\tilde{\theta} TB\log_2(1+\text{SNR}_1 z_1)}\right\} \\ &= -\frac{1}{\tilde{\theta}}\Bigg(\log \mathbb{E}_{z_2}\left\{e^{-\theta_2 TB\log_2(1+\text{SNR}_2 z_2)}\right\} \\ &+ \log \mathbb{E}_{z_1}\left\{e^{(\theta_2-\tilde{\theta}) TB\log_2(1+\text{SNR}_1 z_1)}\right\}\Bigg)\Bigg\}. \end{aligned} \tag{17}$$

**Proof:** We can see from (8) and (9) that

$$\begin{aligned} R &\le -\frac{1}{\tilde{\theta}}\log \mathbb{E}_{z_1}\left\{e^{-\tilde{\theta} TB\log_2(1+\text{SNR}_1 z_1)}\right\} \\ &\le -\frac{1}{\theta_1}\log \mathbb{E}_{z_1}\left\{e^{-\theta_1 TB\log_2(1+\text{SNR}_1 z_1)}\right\}. \end{aligned} \tag{18}$$

**Case I** $\theta_1 \ge \theta_2$:

Consider the link $S-H$. As $\tilde{\theta}$ increases, $R$ decreases. So, $\tilde{\theta} = \theta_1$ returns the highest $R$. Since $\tilde{\theta} \ge \theta_1$, we know that $\tilde{\theta} \ge \theta_2$. If $\hat{\theta} \ge \tilde{\theta}$, the supportable throughput for the link $H-D$ becomes much smaller than necessary when $\hat{\theta} = \theta_2$. As a result, the supported effective bandwidth of the arrivals to $H$ is much smaller, which in turn decreases the effective capacity of the whole system. So, $\hat{\theta} = \theta_2$ returns the highest effective capacity. Considering (12) and the fact that $\tilde{\theta} \ge \hat{\theta}$, we have

$$R \le -\frac{1}{\theta_2}\log \mathbb{E}_{z_2}\left\{e^{-\theta_2 TB\log_2(1+\text{SNR}_2 z_2)}\right\}. \tag{19}$$

Therefore, we find that

$$R_E(\theta_1,\theta_2) = \min\left\{ -\frac{1}{\theta_1}\log \mathbb{E}_{z_1}\left\{e^{-\theta_1 TB\log_2(1+\text{SNR}_1 z_1)}\right\}, -\frac{1}{\theta_2}\log \mathbb{E}_{z_2}\left\{e^{-\theta_2 TB\log_2(1+\text{SNR}_2 z_2)}\right\}\right\}. \tag{20}$$

We can see that for small $\theta_2$, it is possible that the second term inside $\min\{\}$ is greater than the first term, i.e., the effective capacity is unaffected by the intermediate-hop node $H$. This is because of the fact that the effective bandwidth of the departure process from the source can be completely supported by the $H-D$ link when the QoS exponent imposed to the hop node $H$ is small.

If $z_1$ and $z_2$ have the same distribution, we know that

$$\begin{aligned} &-\frac{1}{\theta_2}\log \mathbb{E}_{z_2}\left\{e^{-\theta_2 TB\log_2(1+\text{SNR}_2 z_2)}\right\} \\ &\ge -\frac{1}{\theta_1}\log \mathbb{E}_{z_2}\left\{e^{-\theta_1 TB\log_2(1+\text{SNR}_2 z_2)}\right\} && (21) \\ &\ge -\frac{1}{\theta_1}\log \mathbb{E}_{z_1}\left\{e^{-\theta_1 TB\log_2(1+\text{SNR}_1 z_2)}\right\} && (22) \end{aligned}$$

where (21) and (22) follow from the facts that $-\frac{1}{\theta}\log \mathbb{E}_z\left\{e^{-\theta TB\log_2(1+\text{SNR}z)}\right\}$ is a decreasing function in $\theta$, and a increasing function in SNR. Then, $R_E = -\frac{1}{\theta_1}\log \mathbb{E}_{z_1}\left\{e^{-\theta_1 TB\log_2(1+\text{SNR}_1 z_1)}\right\}$ for all $\theta_2 \le \theta_1$, i.e., the intermediate relay node with the buffer constraints has no negative effect on the effective capacity at all.

**Case II** $\theta_1 < \theta_2$:

According to the previous analysis, we can expect that the effective capacity will be determined by the minimum throughput of the two links $S-H$ and $H-D$. Considering (8) and (11), we can make sure that $\tilde{\theta} \le \hat{\theta}$. If $\tilde{\theta} > \hat{\theta}$, the obtained effective capacity is smaller than the one obtained when $\tilde{\theta} = \hat{\theta}$. Therefore, we must have

$$\Lambda_B(\theta) = R\tilde{\theta} + \log \mathbb{E}_{z_1}\left\{e^{-(\theta-\tilde{\theta})TB\log_2(1+\text{SNR}_1 z_1)}\right\}. \tag{23}$$

1) Suppose that the effective capacity is decided by the link $S-H$ and $\tilde{\theta} = \theta_1$ returns the highest $R$. Combining (11), (12) and (23), we have

$$\begin{aligned} R &\le \sup_{\theta \ge \theta_2} -\frac{1}{\theta_1}\left(\Lambda_H(-\theta) + \Lambda_C(\theta-\theta_1)\right) && (24) \\ &= \sup_{\theta \ge \theta_2} -\frac{1}{\theta_1}\Bigg(\log \mathbb{E}_{z_2}\left\{e^{-\theta TB\log_2(1+\text{SNR}_2 z_2)}\right\} \\ &\quad + \log \mathbb{E}_{z_1}\left\{e^{(\theta-\theta_1) TB\log_2(1+\text{SNR}_1 z_1)}\right\}\Bigg). && (25) \end{aligned}$$

The supremum may not be obtained at $\theta = \theta_2$. This is due to the fact that $\Lambda_H(-\theta)$ is decreasing in $\theta$ while $\Lambda_C(\theta-\theta_1)$ is increasing in $\theta$.

For this case, the necessary condition is

$$
\begin{aligned}
& -\frac{1}{\theta_1}\log\mathbb{E}_{z_1}\left\{e^{-\theta_1 TB\log_2(1+\mathrm{SNR}_1 z_1)}\right\}, \\
& \leq \sup_{\theta\geq\theta_2} -\frac{1}{\theta_1}\bigg(\log\mathbb{E}_{z_2}\left\{e^{-\theta TB\log_2(1+\mathrm{SNR}_2 z_2)}\right\} \\
& \qquad + \log\mathbb{E}_{z_1}\left\{e^{(\theta-\theta_1)TB\log_2(1+\mathrm{SNR}_1 z_1)}\right\}\bigg). \qquad (26)
\end{aligned}
$$

In view of (25), after simple computations, we have

$$
\begin{aligned}
R \leq & \sup_{\theta\geq\theta_2}\frac{\theta}{\theta_1}\Bigg(-\frac{1}{\theta}\log\mathbb{E}_{z_2}\left\{e^{-\theta TB\log_2(1+\mathrm{SNR}_2 z_2)}\right\} \\
& -\left(1-\frac{\theta_1}{\theta}\right)\frac{1}{\theta-\theta_1}\log\mathbb{E}_{z_1}\left\{e^{(\theta-\theta_1)TB\log_2(1+\mathrm{SNR}_1 z_1)}\right\}\Bigg) \\
= & \sup_{\theta\geq\theta_2}\frac{\theta}{\theta_1}\left(E_C(\theta)-E_B(\theta-\theta_1)\right) \qquad (27)
\end{aligned}
$$

where

$$
E_C(\theta) = -\frac{1}{\theta}\log\mathbb{E}_{z_2}\left\{e^{-\theta TB\log_2(1+\mathrm{SNR}_2 z_2)}\right\} \qquad (28)
$$

is the virtual effective capacity with respect to $\theta$, and

$$
E_B(\theta-\theta_1) = \left(1-\frac{\theta_1}{\theta}\right)\frac{1}{\theta-\theta_1}\log\mathbb{E}_{z_1}\left\{e^{(\theta-\theta_1)TB\log_2(1+\mathrm{SNR}_1 z_1)}\right\}
$$

is the virtual effective bandwidth with respect to $\theta-\theta_1$. Similar to the discussion in [5], we know that $E_C(\theta)$ is decreasing in $\theta$ and $E_B(\theta-\theta_1)$ is increasing in $\theta$, and that $E_C(0)$ equals to $\mathbb{E}_{z_2}\{TB\log_2(1+\mathrm{SNR}_2 z_2)\}$ and $E_C(\theta)$ approaches to the delay limited capacity as $\theta\to\infty$, $E_B(\theta-\theta_1)=0$ when $\theta=\theta_1$ and $E_B(\theta-\theta_1)$ approaches to the highest rate of the $S-H$ link as $\theta\to\infty$. Hence, there exists a point $\theta^*$ such that

$$
E_C(\theta^*) = E_B(\theta^*-\theta_1), \qquad (29)
$$

and for $\theta>\theta^*$, $E_C(\theta)<E_B(\theta-\theta_1)$. The right-hand side of (RHS) (25) will have negative values. A numerical result provides a visualization of the above discussion. In Fig. 2, we plot the virtual effective capacity and virtual effective bandwidth normalized by $TB$ as a function of $\theta$ in the Rayleigh fading channel. We assume that $T=2$ ms, $B=10^5$ Hz, $\theta_1=0.01$, $\mathrm{SNR}_1=0$ dB, and $\mathrm{SNR}_2=10$ dB.

Also from the above discussion, we can see that taking the derivative of the RHS of (25) and letting it equal 0, we can solve for the value of $\theta$ where the derivative is 0. If there is one such $\theta^{**}$ in the range $[\theta_2,\theta^*]$, the sup problem is achieved at $\theta^{**}$, otherwise the RHS of (25) is decreasing in $[\theta_2,\theta^*]$ since it is greater than 0 at $\theta=\theta_2$ while equals 0 at $\theta=\theta^*$, and hence sup is obtained at $\theta=\theta_2$. Note here that both $\theta^*$ and $\theta^{**}$ are decided by $\theta_1$, $\mathrm{SNR}_1$ and $\mathrm{SNR}_2$, and are independent of $\theta_2$.

With the above discussion, we can solve for $\theta_2'$ such that the condition (26) is satisfied with equality by removing the sup. For all $\theta_2\leq\theta_2'$, (26) is justified according to (11). This implies that even if the QoS constraint of the intermediate relay node is more stringent than the transmitter, the system can still achieve the same effective capacity as if the QoS constraint of the relay node is small with certain choices of $\mathrm{SNR}_1$ and $\mathrm{SNR}_2$. This is because of the fact that the appropriate choice of $\mathrm{SNR}_2$ can introduce significant increase in the supportable effective bandwidth of the departure process from the source such that it compensates the stringent QoS constraint imposed at the relay node.

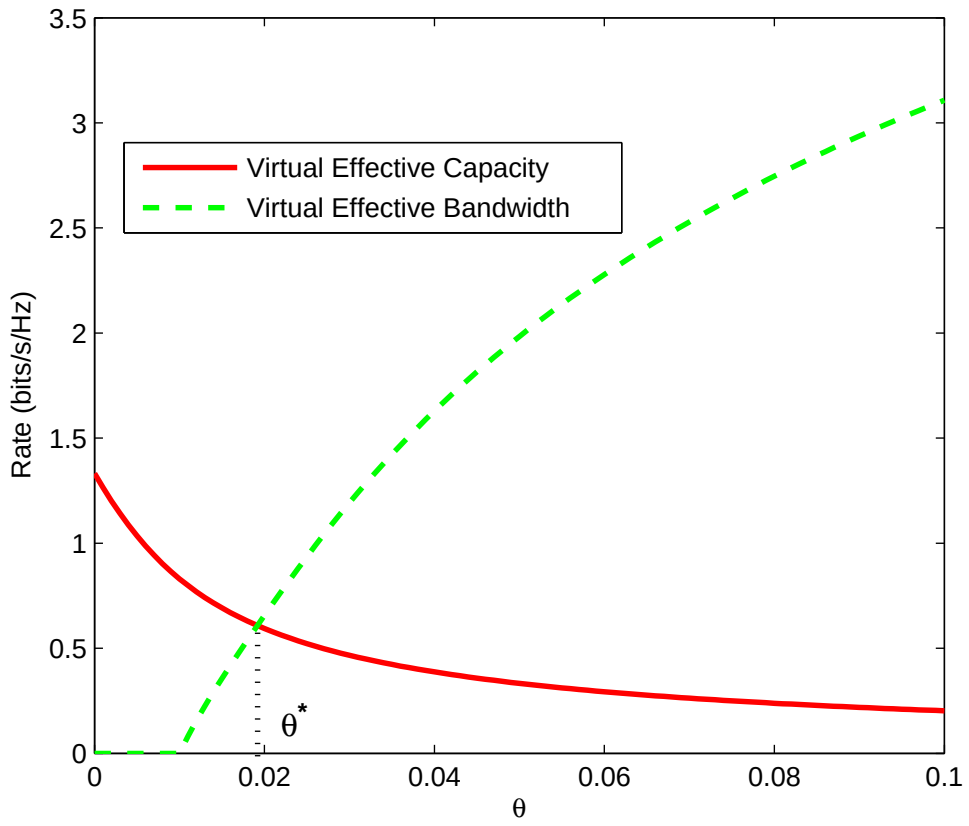

Fig. 2. The virtual effective capacity and virtual effective bandwidth as a function of $\theta$ in Rayleigh fading channels. $\mathbb{E}\{z_1\}=\mathbb{E}\{z_2\}=1$.

2) As shown above, for $\theta_2>\theta_2'$, (26) cannot be satisfied. The effective capacity is not necessarily obtained when $\tilde{\theta}=\theta_1$. Now, $\tilde{\theta}$ can take values from $\theta_1$ to $\hat{\theta}$ to maximize the effective capacity. Note that we can make sure that $\hat{\theta}=\theta_2$. Otherwise, the queue at the relay node is subject to QoS constraints more stringent than necessary, and the effective capacity supported by the link $H-D$ and hence the maximal arrival rate at the transmitter is smaller.

Similar to the above discussions, we have

$$
\begin{aligned}
R \leq & \sup_{\theta_1\leq\tilde{\theta}\leq\theta_2}\min\Bigg\{-\frac{1}{\tilde{\theta}}\log\mathbb{E}_{z_1}\left\{e^{-\tilde{\theta}TB\log_2(1+\mathrm{SNR}_1 z_1)}\right\}, \\
& -\frac{1}{\tilde{\theta}}\left(\Lambda_H(-\theta_2)+\Lambda_C(\theta_2-\tilde{\theta})\right)\Bigg\} \qquad (30) \\
= & \sup_{\theta_1\leq\tilde{\theta}\leq\theta_2}\min\Bigg\{-\frac{1}{\tilde{\theta}}\log\mathbb{E}_{z_1}\left\{e^{-\tilde{\theta}TB\log_2(1+\mathrm{SNR}_1 z_1)}\right\}, \\
& -\frac{1}{\tilde{\theta}}\bigg(\log\mathbb{E}_{z_2}\left\{e^{-\theta_2 TB\log_2(1+\mathrm{SNR}_2 z_2)}\right\} \\
& + \log\mathbb{E}_{z_1}\left\{e^{(\theta_2-\tilde{\theta})TB\log_2(1+\mathrm{SNR}_1 z_1)}\right\}\bigg)\Bigg\}. \qquad (31)
\end{aligned}
$$

Obviously, the first term inside $\min\{\}$ is decreasing in $\tilde{\theta}$, while the second term is increasing in $\tilde{\theta}$. Hence, the point that

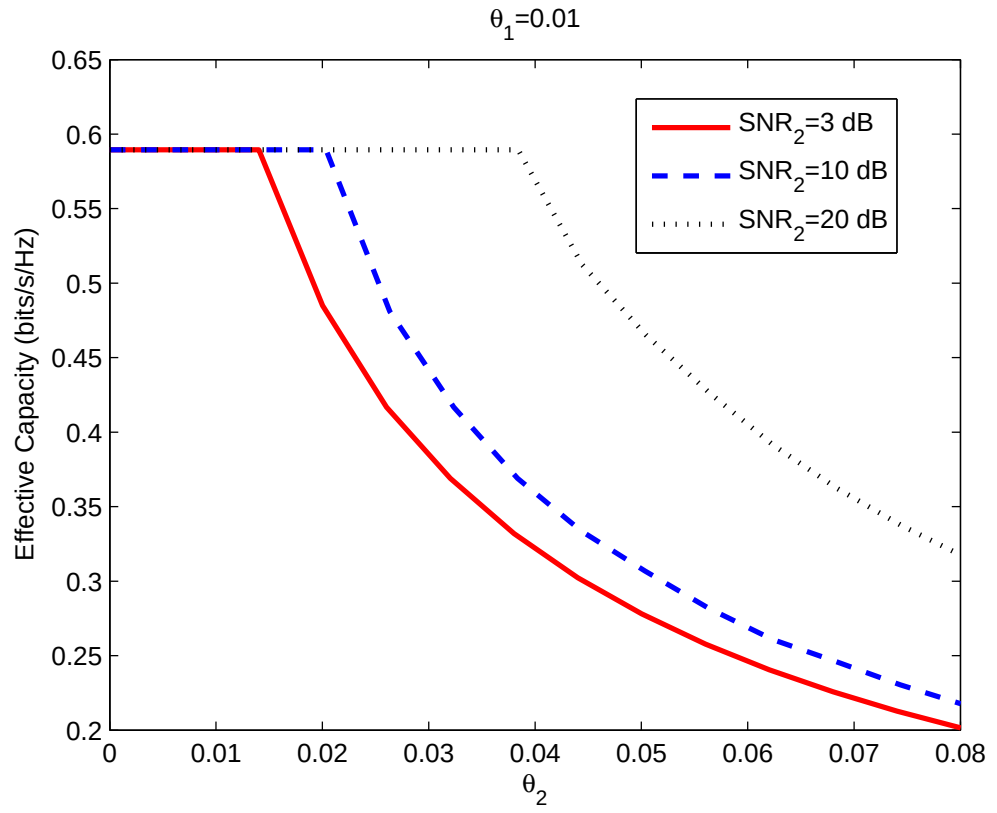

Fig. 3. The effective capacity as a function of $\theta$ in Rayleigh fading channels. $\mathbb{E}\{z_1\} = \mathbb{E}\{z_2\} = 1$.

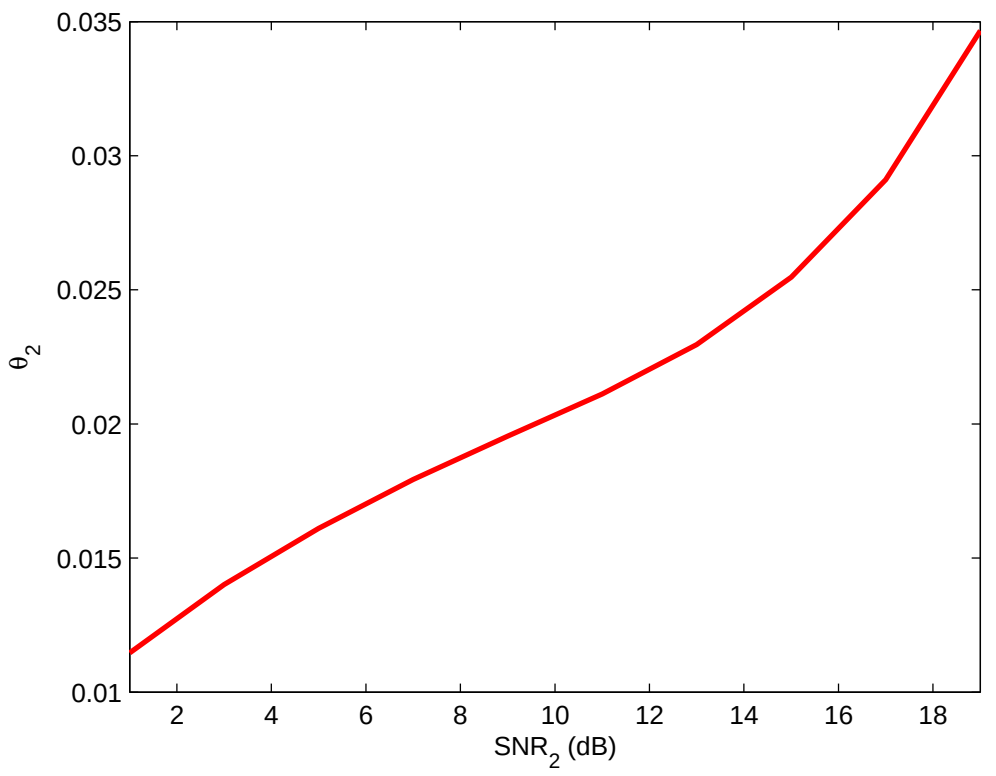

Fig. 4. $\theta_2'$ vs. $\text{SNR}_2$ in Rayleigh fading channels. $\mathbb{E}\{z_1\} = \mathbb{E}\{z_2\} = 1$.

maximizes the value is obtained when

$$-\frac{1}{\tilde{\theta}}\log \mathbb{E}_{z_1}\left\{e^{-\tilde{\theta}TB\log_2(1+\text{SNR}_1 z_1)}\right\}$$
$$= -\frac{1}{\tilde{\theta}}\left(\log \mathbb{E}_{z_2}\left\{e^{-\theta_2 TB\log_2(1+\text{SNR}_2 z_2)}\right\}\right.$$
$$\left.+\log \mathbb{E}_{z_1}\left\{e^{(\theta_2-\tilde{\theta})TB\log_2(1+\text{SNR}_1 z_1)}\right\}\right)\Bigg\}. \tag{32}$$

Note that as long as the highest rate of the $S-H$ link is higher than the delay limited capacity of the $H-D$ link , we should have $\tilde{\theta} \to \theta_2$ as $\theta_2 \to \infty$ from the previous discussion about virtual effective capacity and virtual effective bandwidth. Otherwise, (31) turns out to be negative valued, and as a result $R_E$ becomes 0. So, for the special case of $\theta_2 \to \infty$, we only need to compare the delay limited capacity of the two links $S-H$ and $H-D$, and choose the smallest one as the effective capacity. □

We assume $\theta_1 = 0.01$ in the following numerical results. In Fig. 3, we plot the effective capacity as a function of the QoS constraints of the intermediate relay node for different $\text{SNR}_2$ values. From the figure, we can see that the effective capacity does not decrease for a certain range of $\theta_2$, and this range is increased by $\text{SNR}_2$. Motivated by this observation, we plot the value of $\theta_2'$, up to which the effective capacity is unaffected, as a function of $\text{SNR}_2$ in Fig. 4. Note that for all values of $(\text{SNR}, \theta_2)$ below the curve shown in the figure, the QoS constraints of the relay node do not impose any negative effect on the effective capacity. This provides us with useful insight on the design of wireless systems.

## IV. Conclusion

In this paper, we have analyzed the maximum arrival rates that can be supported by a two-hop communication link in which the source and relay nodes are subject to buffer constraints. We have determined the effective capacity in the block-fading scenario as a function of the signal-to-noise ratio levels $\text{SNR}_1$ and $\text{SNR}_2$ and the QoS exponents $\theta_1$ and $\theta_2$. We have found that when the QoS exponent $\theta_2$ of the relay node is small, the effective capacity is not affected by the buffer constraints at the relay. We have shown that as the SNR level at the relay node increases, the effective capacity can stay unaffected for larger values of the QoS exponent imposed at the relay node. Also, as $\theta_2$ increases, effective capacity approaches the minimum of the delay limited capacity of the two links.